\documentclass[letterpaper, 10 pt, conference]{ieeeconf}  

\IEEEoverridecommandlockouts                              

\usepackage{epsfig}
\usepackage{amsmath}
\usepackage{cleveref}
\usepackage{footmisc}
\usepackage{color}
\usepackage{multirow}
\usepackage{textcomp}
\usepackage{amssymb}

\title{Z-Net: an Anisotropic 3D DCNN for Medical CT Volume Segmentation
}

\author{Peichao Li$^{1*}$ and Xiao-Yun Zhou$^{1*}$ and Zhao-Yang Wang$^{1}$ and Guang-Zhong Yang$^{1,2}$
\thanks{*Peichao Li and Xiao-Yun Zhou contribute equally to this paper}
\thanks{$^{1}$The Hamlyn Centre for Robotic Surgery, Imperial College London, UK
        {\tt\small llhpqlcsg@gmail.com, xiaoyun.zhou14@imperial.ac.uk}}%
\thanks{$^{2}$Institute of Medical Robotics, Shanghai Jiao Tong University, China}%
}

\begin{document}

\maketitle
\thispagestyle{empty}
\pagestyle{empty}

\begin{abstract}
Accurate volume segmentation from the Computed Tomography (CT) scan is a common prerequisite for pre-operative planning, intra-operative guidance and quantitative assessment of therapeutic outcomes in robot-assisted Minimally Invasive Surgery (MIS). 3D Deep Convolutional Neural Network (DCNN) is a viable solution for this task, but is memory intensive. Small isotropic patches are cropped from the original and large CT volume to mitigate this issue in practice, but it may cause discontinuities between the adjacent patches and severe class-imbalances within individual sub-volumes. This paper presents a new 3D DCNN framework, namely Z-Net, to tackle the discontinuity and class-imbalance issue by preserving a full field-of-view of the objects in the XY planes using anisotropic spatial separable convolutions. The proposed Z-Net can be seamlessly integrated into existing 3D DCNNs with isotropic convolutions such as 3D U-Net and V-Net, with improved volume segmentation Intersection over Union (IoU) - up to $12.6\%$. Detailed validation of Z-Net is provided for CT aortic, liver and lung segmentation, demonstrating the effectiveness and practical value of Z-Net for intra-operative 3D navigation in robot-assisted MIS. 
\end{abstract}

\section{Introduction}
\label{sec: Intro}
\begin{figure}
    \centering
    \framebox{\includegraphics[width=0.25\textwidth]{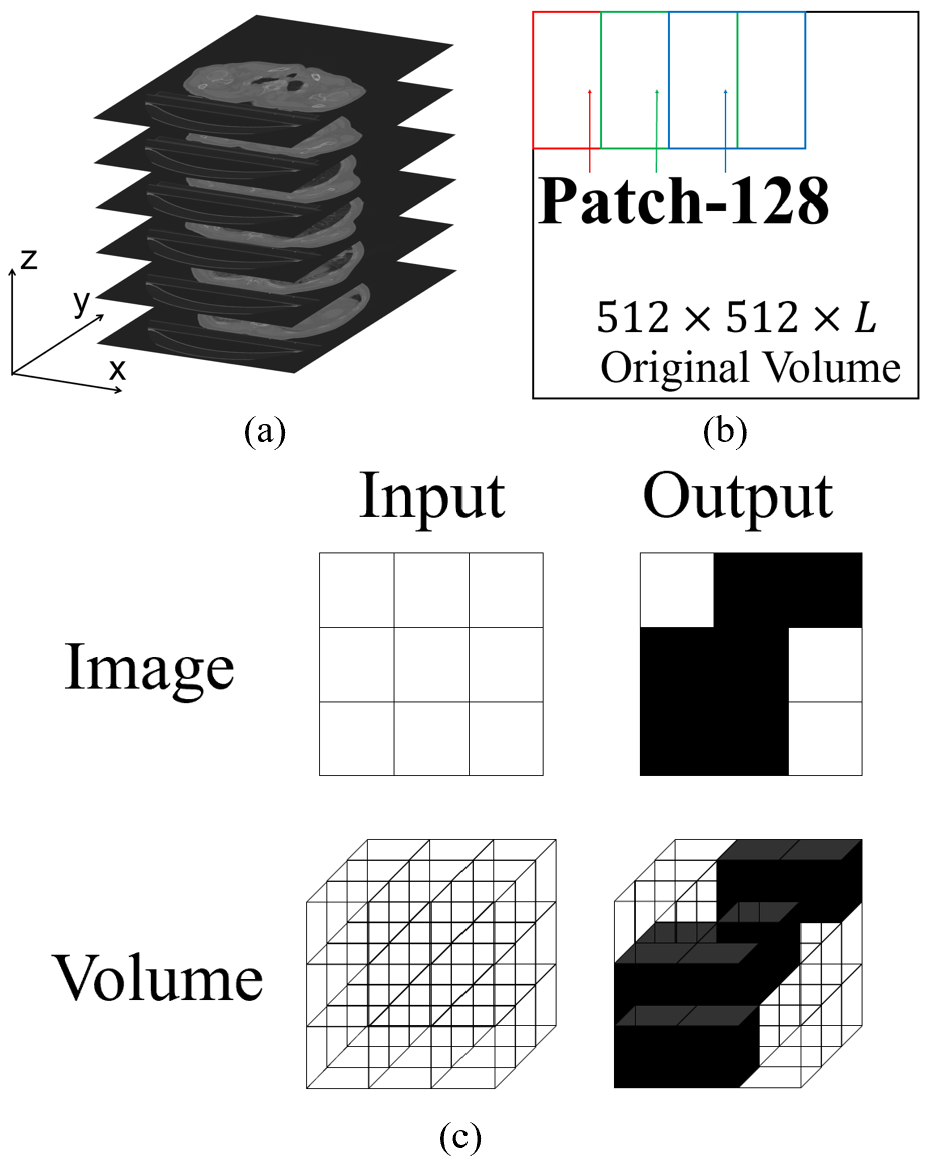}}
    \caption{Illustrations of (a) CT volume data and the definition of X, Y and Z axes; (b) Patch-128 which crops the original CT volume of size $512\times 512 \times \rm L$ to multiple small patches of size $128\times 128 \times 64$; (c) comparison between image segmentation and volume segmentation.}
    \label{fig:intro}
\end{figure}

Medical volume segmentation, which labels the class of each voxel in a 3D volume, including the anatomy, prosthesis and lesion, is an important task in medical image analysis. Common 3D medical volume acquisition techniques include Computed Tomography (CT), Magnetic Resonance Imaging (MRI), 3D ultrasound and so on. MRI can achieve good image quality without radiation, but at the cost of longer imaging time. 3D ultrasound can achieve real-time 3D imaging, but with less-than-optimal image quality. CT is able to retrieve detailed and high-resolution volumetric representations of human structures and records them as voxel values, at the cost of radiation. It is widely used for pre-operative diagnosis, intra-operative navigation and post-operative assessment. In this paper, we mainly focus on medical CT volume segmentation. An illustration of a 3D CT volume and the definition of the three dimensions used in this paper is shown in Fig. \ref{fig:intro}(a).

Traditionally, in open surgeries operated through a large incision larger than $10cm$, CT volume segmentation was mostly used for pre-operative diagnosis and post-operative assessment. For example, the liver and tumour were segmented by Hybrid Densely Connected UNet (H-DenseUNet) from CT volumes to diagnose the hepatocellular carcinoma \cite{li2018h}. The abdominal aortic thrombus was segmented from CT volumes to assess the Endovascular Aneurysm Repair (EVAR) operation outcomes for treating Abdominal Aortic Aneurysm (AAA) \cite{lopez2018fully}.

Recently, due to the emerging of robot-assisted Minimally Invasive Surgeries (MISs) where intelligent robotic surgical tools are inserted through a small incision less than $2cm$ \cite{vitiello2012emerging}, i.e. Laparo-Endo-Scopic Single-site (LESS) surgery, or a natural orifice, i.e. Natural Orifice Transluminal Endoscopic Surgery (NOTES) \cite{liu2018design}, CT volume segmentation of organs and prostheses is becoming increasingly helpful in intra-operative surgical robotic navigation. For example, the 3D Right Ventricle (RV) mesh that is segmented from pre-operative CT volume was an essential input for the mapping vertex determination and hence efficient robotic path planning in robot-assisted Radio-frequency Cardiac Ablation (RFCA) \cite{zhou2016path}. The 3D model of the aorta segmented from the pre-operative CT volume was essential to instantiate a safe robotic path for navigating Fenestrated Endovascular Aneurysm Repair (FEVAR) \cite{zheng2019towards}. The 3D liver segmentation was used to instantiate the intra-operative and instantaneous 3D liver shapes for navigating robot-assisted liver surgeries \cite{zhou2018real}. Apart from organ segmentation, prosthesis segmentation is also very useful in robot-assisted MISs. For example, the 3D stent graft and 3D marker shape segmented from the pre-operative CT volume were used to instantiate the intra-operative 3D shape of a fully-compressed \cite{zhoustent}, partially-deployed \cite{zheng2019real} and fully-deployed \cite{zhou2018real-instantiation} fenestrated stent graft, improving the navigation for FEVAR from 2D to 3D.

Volume segmentation can be realized by either stacking the results of multiple 2D image segmentation, which labels the class that each pixel in a 2D image belongs to, or by direct volume segmentation, which considers the inter-slice connection and information between voxels in 3D. An illustration of the difference between image and volume segmentation is shown in Fig. \ref{fig:intro}(c). Since the success of AlexNet \cite{krizhevsky2012imagenet}, Deep Convolutional Neural Network (DCNN) has been widely used to replace traditional hand-crafted feature extractors, i.e. edge detector with filters, as it can achieve automatic feature extraction and pixel probability regression with an end-to-end training manner. 3D DCNN has also been a popular technique for direct 3D volume segmentation. In image segmentation, all operations including convolutional layers, max-pooling layers and transposed convolutional layers are in 2D, while in volume segmentation, all these operations are in 3D.

For example, 3D U-Net was first proposed with a contracting encoder part, an expanding decoder part and long skip connections for kidney volume segmentation \cite{cciccek20163d}. Similar to 3D U-Net architecture, V-Net was introduced with a larger $5 \times 5 \times 5$ convolutional kernel and residual learning for prostate volume segmentation \cite{milletari2016v}. 3D Deeply Supervised Network (DSN) was proposed to demonstrate the effect of both supervising the lower and upper DCNN layers for liver and cardiac volume segmentation \cite{dou20173d}. Multi-scale DCNN and Conditional Random Field (CRF) were combined for brain lesion volume segmentation \cite{kamnitsas2017efficient}. Holistic Decomposition Convolution (HDC) was proposed to use a larger size of input through decomposed convolutions, resulting in an improved accuracy for medical volume segmentation \cite{zeng2019holistic}.

In most previous works, features and convolutional kernels in 3D DCNNs are usually symmetric in all dimensions, where the calculation along the Z axis is similar to or the same as that along the XY axes. Compared to 2D DCNN, this additional calculation along the Z axis increases the memory usage on the Graphics Processing Unit (GPU). Due to the memory limitation of current GPU capacity, it is usually impossible to hold an entire CT volume of a patient, typically with a size of $512 \times 512 \times \rm L$, $\rm L > 400$, as a training volume input to a 3D DCNN. A popular solution is by patch division, which refers to cropping the original medical CT volume into small patches either randomly or selectively. For example, 3D U-Net \cite{cciccek20163d} cropped the original and large CT volume into $132\times132\times116$ patches as the network input. V-Net \cite{milletari2016v} cropped with a size of $128\times128\times64$. A size of $160 \times 160 \times 72$ was used in DSN \cite{dou20173d} while sizes of $25 \times 25 \times 25$ and $19 \times 19 \times 19$ were used in multi-scale DCNN \cite{kamnitsas2017efficient}. An illustration of cropping the sub-volume patches from the original CT volume is shown in Fig. \ref{fig:intro}(b). In addition to solve the GPU memory limitation, patch division also helps with the limitation of training data. Smaller patches allow the augmentation of one training volume into multiple training patches, current public dataset is usually of less than 100 training volumes and this number is insufficient for training 3D DCNNs. However, deficiencies exist for current patch division methods with small and nearly-isotropic patch sizes: 1) class-imbalance: some patches might contain very few or no foreground, while some others may have the entire sub-volume as foreground; 2) limited field-of-view: each patch can only perceive a small portion of the entire volume of a patient, and thus is not able to extract a global context. 3) discontinuous segmentation results: the segmentation of the entire CT volume is stacked together from the results of sub-volume patches, which may introduce discontinuities at the boundaries.

In this paper, in order to deal with the three deficiencies in 3D DCNN with isotropic convolutions, we propose a novel 3D DCNN framework - Z-Net, which processes the features in the XY-plane and the features in the Z-plane separately with anisotropic spatial separable convolutions. The motivation of Z-Net is that inter-slice information is helpful for 3D segmentation, but it may not need a large amount of slices in the Z axis, i.e. 116 in 3D U-Net and 64 in V-Net. With an independent calculation along the Z axis separated from the calculation along the XY axes using isotropic spatial separable convolution, four advantages can be introduced:
\begin{itemize}
    \item A full field-of-view along the axial slices (XY-plane) is maintained
    \item Segmentation results are continuous in the axial slices (XY-plane)
    \item Less class-imbalance exists as the axial slices (XY-plane) contain both the foreground and background
    \item Sufficient training data can be augmented as only 8 rather than $\rm L$ slices are needed in one patch
    \item Inter-slice information across up to 8 slices can be extracted;
    \item Reduced trainable parameters, calculation amount and training time
\end{itemize}

The spatial separable convolution technique is adopted in Z-Net to divide a high dimensional kernel into several lower dimensional kernels to reduce the amount of computation. For example, a $3 \times 3$ kernel can be divided into a $3\times1$ kernel and a $1\times3$ kernel. Therefore, the number of multiplications is reduced from 9 to 6. Early separable convolution can be traced back to 2012 when Mamalet \textit{et al.} proposed several methods to simplify the filters in convolutional networks \cite{mamalet2012simplifying}. Sironi \textit{et al.} further analyzed the separable convolution mathematically, and proposed to use tensor decomposition to get a basis of separable filters for approximation of the high-rank kernels \cite{sironi2014learning}. It was demonstrated that the decomposed convolutions derived from low-rank approximation can reduce the computational complexity without significant changes in the accuracy. Peng \textit{et al.} adopted the separable convolutions in their network design for semantic segmentation \cite{peng2017large}. It argued that contrary to image classification tasks where smaller kernels and deeper networks might be ideal, segmentation accuracy may benefit from larger kernels. A global convolutional network with massive convolutional kernels was proposed, and spatial separable filters were used to reduce the computational cost.

The proposed Z-Net is a framework that processes the features in XY-planes and the features in Z-plane separately. It can be seamlessly integrated to the current popular medical volume segmentation DCNNs, i.e. 3D U-Net or V-Net, resulting in ZU-Net and ZV-net respectively. The main contributions of this paper are:
\begin{itemize}
    \item A new 3D DCNN framework - Z-Net is proposed for medical volume segmentation. It involves separating traditional 3D convolutional kernels into combinations of 2D and 1D convolutional kernels. It can be seamlessly integrated to current and popular 3D DCNNs for CT volume segmentation. In addition, the proposed Z-Net decreases the trainable parameters and training time significantly.
    \item Z-Net addresses the problems of the class-imbalance, limited field-of-view and discontinuous segmentation results in the existing 3D DCNN with isotropic shapes of features and kernels.
    \item Aortic, liver and lung CT volumes are used as the validation with detailed ablation analysis. The proposed Z-Net can achieve an improvement in segmentation accuracy of up to $12.6\%$ compared to 3D U-Net, in terms of the Intersection over Union (IoU).
\end{itemize}

The proposed Z-Net and experimental setup are introduced in Sec. \ref{sec:method}. Detailed validations comparing the proposed Z-Net to baselines, also segmentation examples and the training loss curve are stated in Sec. \ref{sec:result}. Discussion and conclusion are shown in Sec. \ref{sec:discussion} and Sec. \ref{sec:conclusion} respectively.

\section{Methodology}
\label{sec:method}

\subsection{Z-Net}
\label{sec:znet}
Z-Net is a framework that can be seamlessly integrated to existing 3D DCNNs with isotropic shapes of features and kernels. In this paper, we will firstly introduce two typical examples - 3D U-Net and V-Net, and then provide detailed information about the proposed Z-Net framework.

\subsubsection{Traditional 3D U-Net and V-Net}

The input feature map for a 3D DCNN can be represented as $\boldsymbol{\rm F} \in \mathbb{R}^{\rm N \times H \times W \times D \times C_{in}}$, where $\rm N$ is the batch size, $\rm H$ is the height, $\rm W$ is the width, $\rm D$ is the depth and $C_{in}$ is the number of channels. Unlike natural RGB images with three channels, CT scan is with a single channel, hence $C_{in} = 1$ when this feature map represents the input patch.

3D convolution uses a trainable 3D convolutional kernel $\boldsymbol{\rm T} \in \mathbb{R}^{\rm H_T \times W_T \times D_T}$ which slides through the height, width and depth of the input with a stride of $\rm S$ to calculate the convolution. This operation can be represented as:

\begin{equation}
\boldsymbol{\hat{\rm F}} = \phi(\boldsymbol{\rm F}*\boldsymbol{\rm T}+b)
\end{equation}
where $\boldsymbol{\hat{\rm F}} \in \mathbb{R}^{\rm N \times H' \times W' \times D' \times C_{out}}$ represents the output feature map, $b$ is the bias, $\phi(\cdot)$ represents the activation function to add non-linearity to the networks, typically Sigmoid function or Rectified Linear Unit (ReLU) function. $C_{out}$ is the number of output feature channels, $\rm H'=H//S$, $\rm W'=W//S$, $\rm D'=D//S$ where $//$ represents the floor division. When $\rm S=1$, the operation is equivalent to a normal convolution, resulting in feature maps with the same dimensional size of the input feature map, provided that a proper padding method has been adopted. When $\rm S>1$, the convolutional operation generates a down-sampled feature map. When $\rm S<1$, the convolutional operation generates an up-sampled feature map, which is named transposed convolution. Another commonly used down-sampling operation is max-pooling, where the maximum values within the $\rm H_T \times W_T \times D_T$ region are extracted to represent the area. After the convolutional operation, $\boldsymbol{\hat{\rm F}}$ is then passed into the normalization layer, where the mean and variance are calculated as:

\begin{equation}
\mu_{n,c} = \frac{1}{\rm H \times \rm W \times \rm D} \sum\limits_{h=1}^{\rm H}  \sum\limits_{d=1}^{\rm D} \sum\limits_{w=1}^{\rm W} \hat{f}_{n,h,w,d,c}
\end{equation}

\begin{equation}
\delta_{n,c}^2 = \frac{1}{\rm H \times \rm W \times \rm D} \sum\limits_{h=1}^{\rm H} \sum\limits_{w=1}^{\rm W} \sum\limits_{d=1}^{\rm D} (\hat{f}_{n,h,w,d,c}-\mu_{n,c})^2
\end{equation}
where $\hat{f}_{n,h,w,d,c}$ represents each individual voxel value inside the $\boldsymbol{\hat{\rm F}}$. All data inside the feature map is normalized to a mean of 0 and a variance of 1 to facilitate the training, then is re-scaled by $\gamma_{n,c}$ and re-translated by $\beta_{n,c}$ to maintain the DCNN representation ability:

\begin{equation}
\hat{f'}_{n,h,w,d,c} = \frac{\hat{f}_{n,h,w,d,c}-\mu_{n,c}}{\sqrt{\delta_{n,c}^2+\epsilon}} \times \gamma_{n,c} + \beta_{n,c}
\end{equation}

\begin{figure}
    \centering
    \framebox{\includegraphics[width=0.45\textwidth]{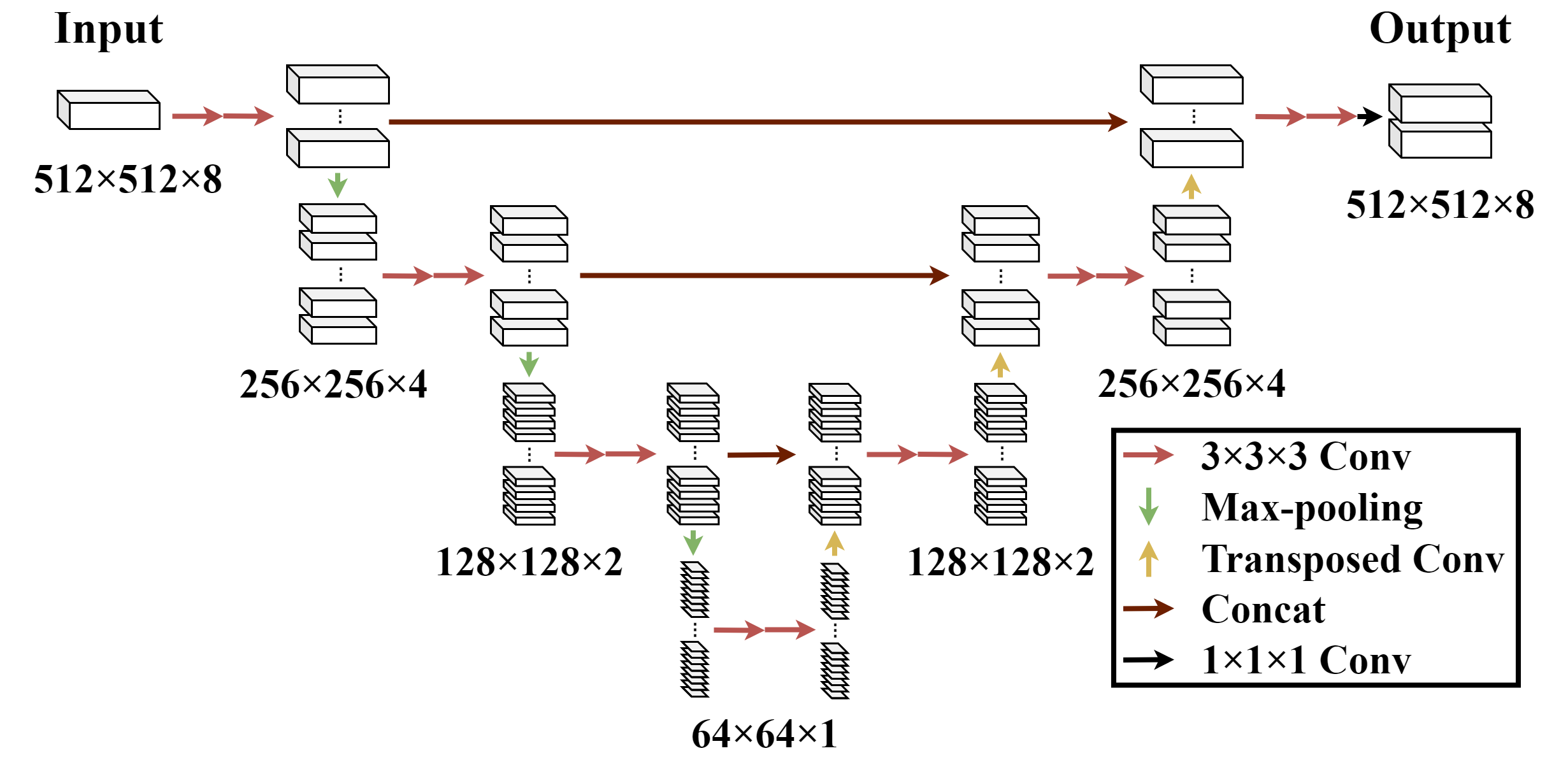}}
    \caption{The detailed network architecture of 3D U-Net}
    \label{fig:unet}
\end{figure}

Most 3D DCNNs consist of several convolutional, max-pooling, transposed convolutional, normalization and ReLU layers. Typical examples are 3D U-Net \cite{cciccek20163d} and V-Net \cite{milletari2016v}. In 3D U-Net, as illustrated in Fig. \ref{fig:unet}, the contracting encoder part, which contains 3D convolutional layers and 3D max-pooling layers, is used to down-sample the input patch to feature maps in different resolutions, while the expanding decoder part with 3D convolutional layers and 3D transposed convolutional layers is used to recover the feature maps until they reach the original resolution. Skip connections are used to concatenate feature maps from the contracting path to the expanding path to facilitate information propagation. All convolutional layers are with a kernel size of 3 and max-pooling layers are with a kernel size/stride of 2. Two 3D convolutional layers are used before each 3D max-pooling or 3D transposed convolutional layers.

\begin{figure}
    \centering
    \framebox{\includegraphics[width=0.45\textwidth]{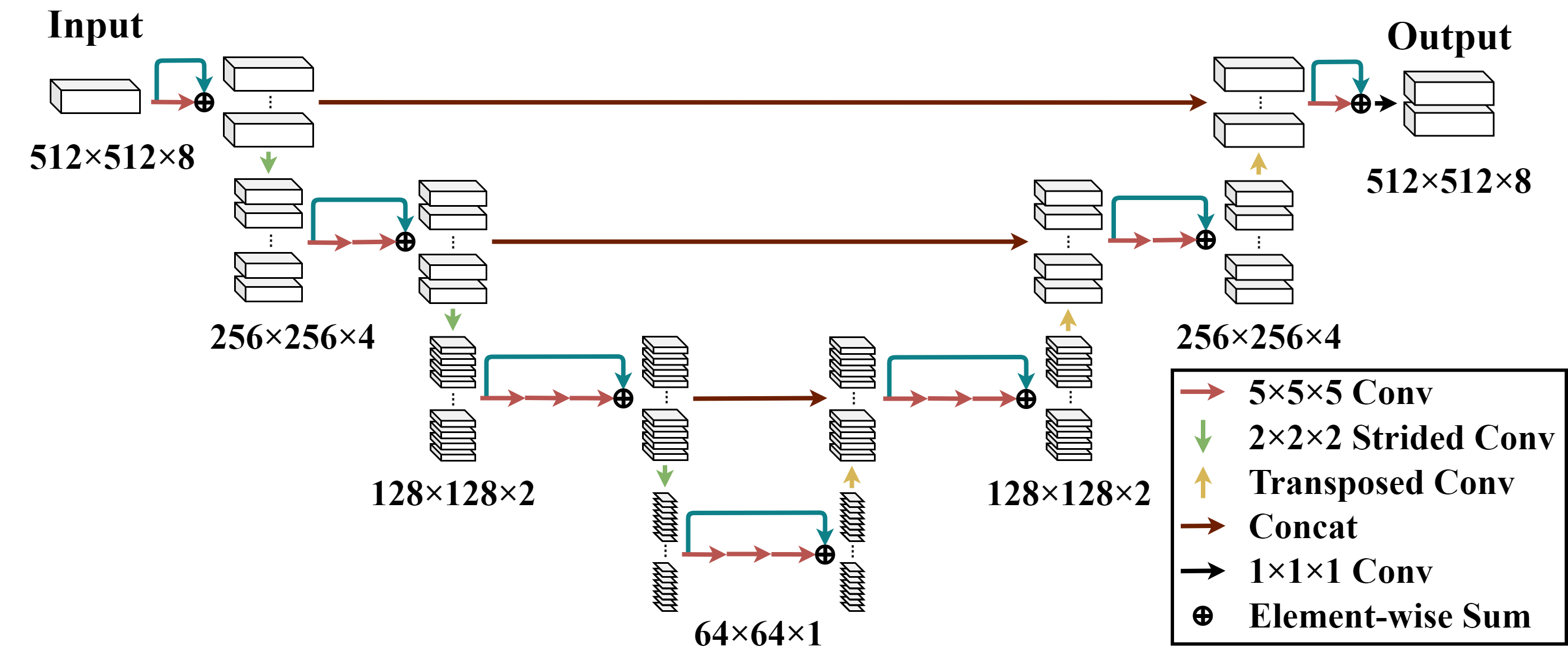}}
    \caption{The detailed network architecture of V-Net}
    \label{fig:vnet}
\end{figure}

The network architecture of V-Net, which is shown in Fig. \ref{fig:vnet}, is similar to 3D U-Net. Residual connections are used to add feature maps from shallow layers to feature maps from deep layers to facilitate the training, similar to residual learning \cite{he2016deep}. Kernel size of 5 is used for all convolutional layers. Convolutional layers with a stride of 2 and transposed convolutional layers are used as the down-sampling and up-sampling layers respectively. The number of convolutional layers before down-sampling and up-sampling layers increases in the contracting path, but it decreases in the expanding path.

\subsubsection{Z-Net}

\begin{figure}
    \centering
    \includegraphics[width=0.25\textwidth]{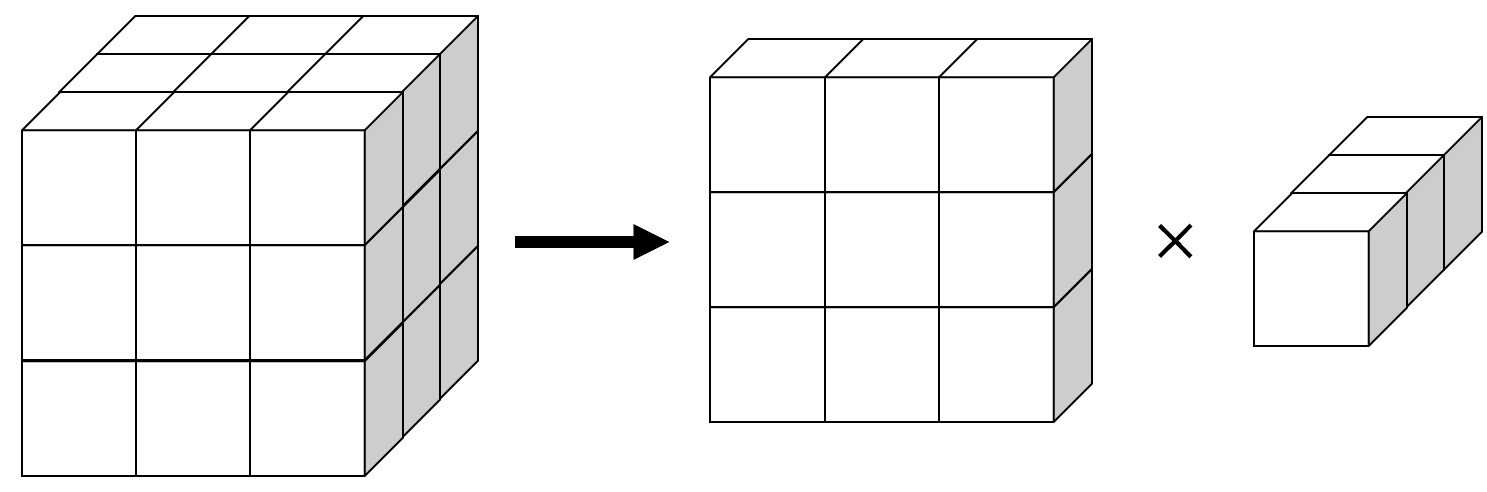}
    \caption{In anisotropic spatial separable convolutions, a $3\times3\times3$ convolutional filter can be decomposed into a 2D $3\times3\times1$ filter and a 1D $3\times3\times1$ filter.}
    \label{fig:separable_conv}
\end{figure}

The proposed Z-Net is a framework for designing or modifying a 3D DCNN architecture for CT volume segmentation, where traditional 3D convolutional operations are decomposed into 2D convolutions along XY-plane and a 1D convolution along Z axis. For example, a 3D convolutional kernel of size $3 \times 3 \times 3$ is separated into a 2D convolutional kernel of size $3 \times 3 \times 1$ and a 1D convolutional kernel of size $1 \times 1 \times \rm D$, where $\rm D$ is the depth of the input feature map. The kernel size of 1D convolutional kernel is set to $1 \times 1 \times \rm D$ rather than $1 \times 1 \times 3$ to fully extract the inter-slice context among all slices without significantly increase the computational cost. Such decomposition is illustrated in Fig. \ref{fig:separable_conv}.

The 2D convolution applies a trainable 2D convolutional kernel $\boldsymbol{\rm T_2} \in \mathbb{R}^{H_T \times W_T \times 1}$ that sums the multiplications along the height, width, and depth of the input volume:

\begin{equation}
\boldsymbol{\hat{\rm F}} = \phi(\boldsymbol{\rm F} * \boldsymbol{\rm T_2}+b)
\end{equation}

The 1D convolution uses a 1D convolutional kernel $\boldsymbol{\rm T_1} \in \mathbb{R}^{1 \times 1 \times D_T}$ that moves along the height, width, and depth of the input:

\begin{equation}
\boldsymbol{\hat{\rm F}} = \phi(\boldsymbol{\rm F} * \boldsymbol{\rm T_1}+b)
\end{equation}

\begin{figure}
    \centering
    \framebox{\includegraphics[width=0.45\textwidth]{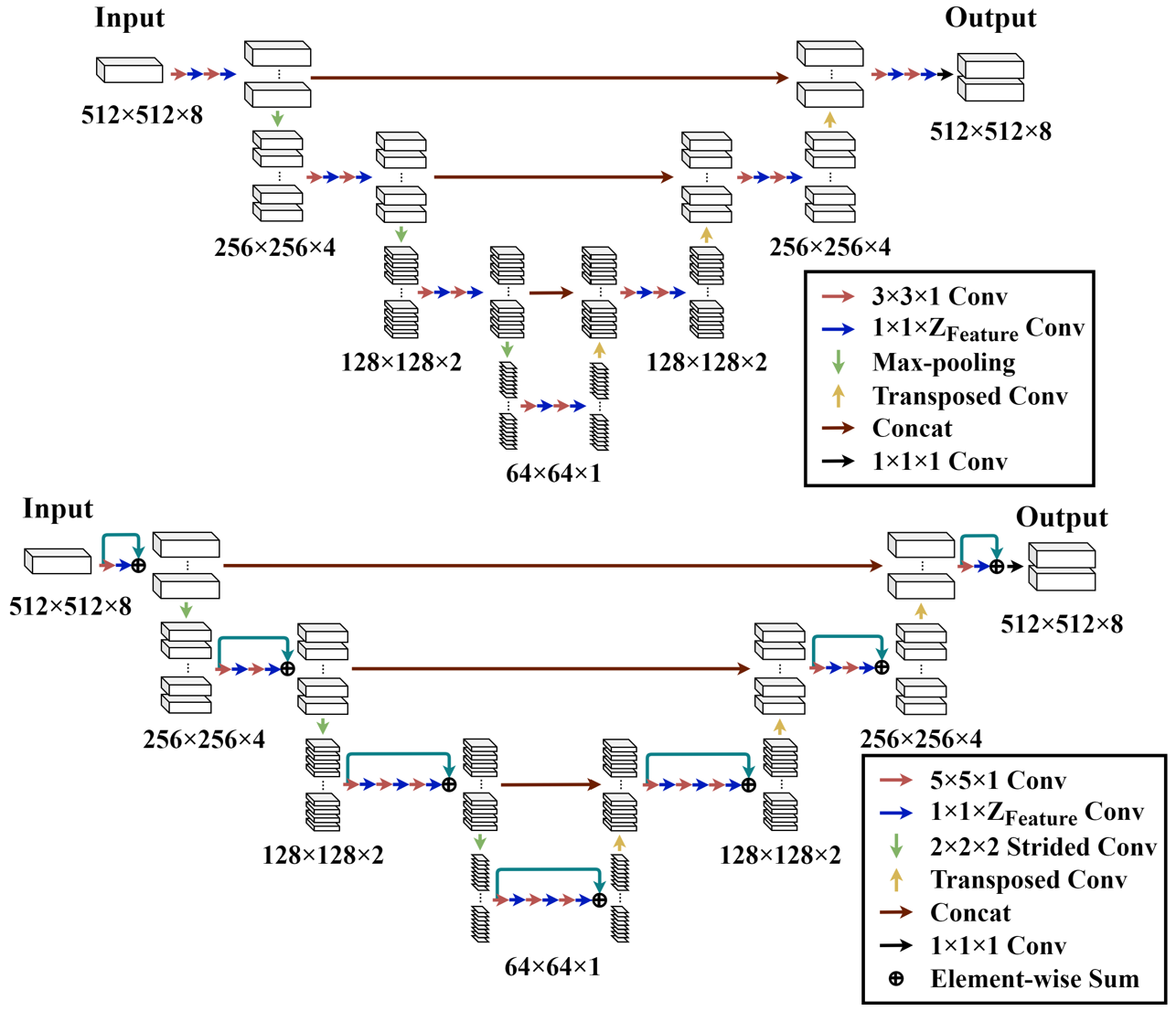}}
    \caption{The detailed network architecture of ZU-Net v1 (top) and ZV-Net v1 (bottom) under mode 1.}
    \label{fig:mode1}
\end{figure}

\begin{figure}
    \centering
    \framebox{\includegraphics[width=0.45\textwidth]{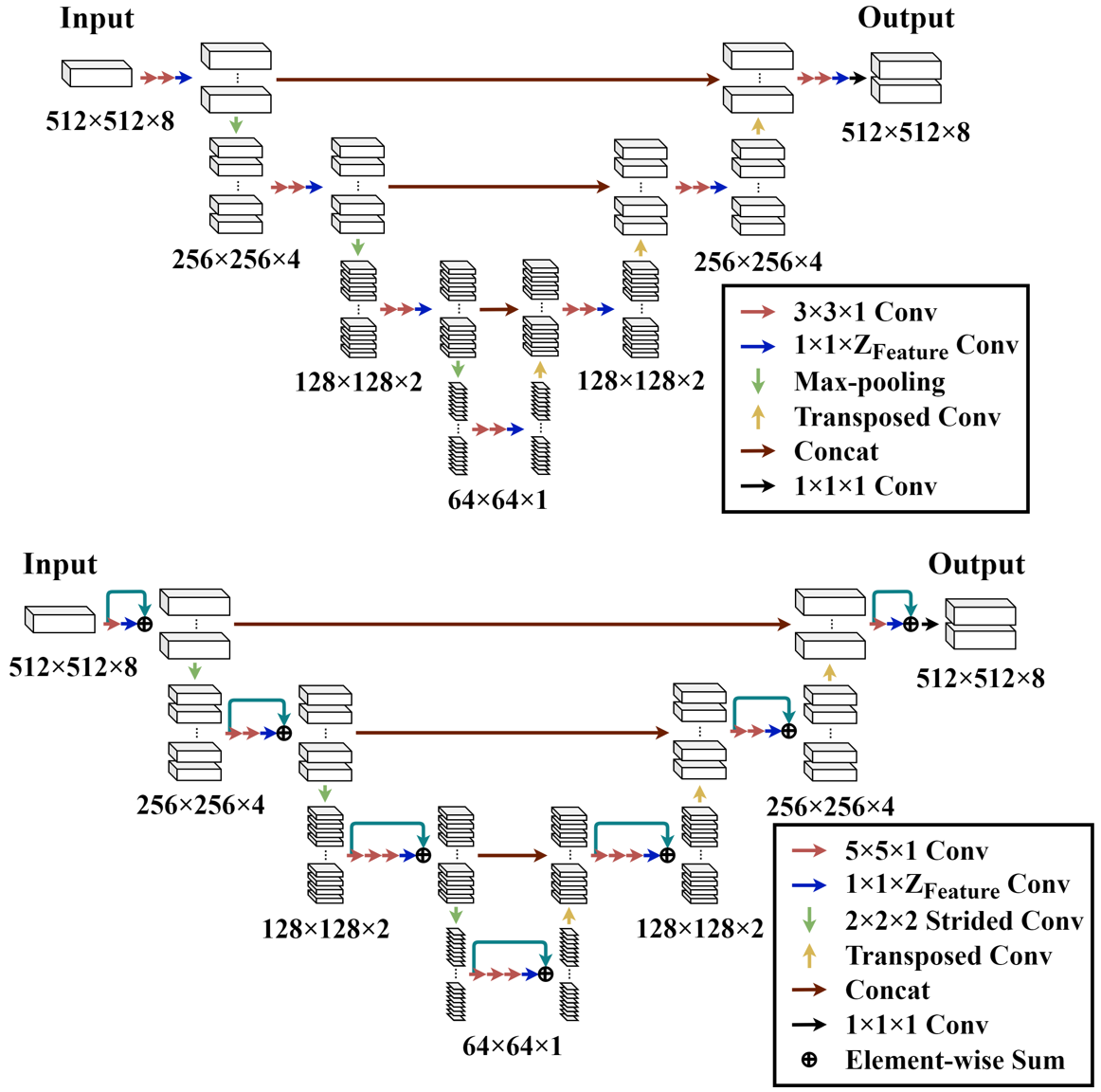}}
    \caption{The detailed network architecture of ZU-Net v2 (top) and ZV-Net v2 (bottom) under mode 2.}
    \label{fig:mode2}
\end{figure}

In Z-Net, except the down-sampling and up-sampling layers, all 3D convolutional layers are replaced with spatial separable convolutions. It is easy to integrate the proposed Z-Net to traditional 3D DCNNs. In this paper, we take two typical examples - 3D U-Net and V-Net architectures for illustration. Two versions of the modified DCNN architectures for each network are explored: 1) mode 1: replace all 3D convolutional layers in 3D U-Net and V-Net with 2D convolutional layers followed by 1D convolutional layers. These changes are reflected by ZU-Net v1 and ZV-Net v1, as shown in Fig. \ref{fig:mode1}; 2) mode 2: replace all 3D convolutional layers in 3D U-Net and V-Net with 2D convolutional layers, and only add a single 1D convolutional layer before each down-sampling or up-sampling layer, which means all intermediate 1D convolutions in ZU-Net v1 and ZV-Net v1 are removed. The network architectures, namely ZU-Net v2 and ZV-Net v2, are shown in Fig. \ref{fig:mode2}. Given that Instance Normalization (IN) outperforms other normalization methods including Batch Normalization (BN), Layer Normalization (LN), Group Normalization (GN) as proved in \cite{zhou2019normalization}, 3D IN was used in this paper for all DCNNs as the default normalization method.

A typical volume size for CT scan is $512 \times 512 \times \rm L$, where $\rm L$ is the number of slices along the Z axis which is varied for different subjects, typically larger than 400. In traditional 3D DCNNs, Patch-128 and Patch-64 are common methods for cropping the original CT volume, which represents the patch size of $128 \times 128 \times 64$ and $64 \times 64 \times 64$ respectively. With the proposed Z-Net, the original CT volume is cropped into sub-volumes of size of $512 \times 512 \times 8$ (named Patch-512), and the stride between the successive crops is 1, which leads to $\rm (L-1)$ augmented training patches from each patient's CT volume. 

Z-Net maintains a full field-of-view in the XY slices while it becomes smaller along the Z axis to feed into a single GPU. As sub-volume divisions will limit the effective field-of-view for the network to perceive the entire volume of a subject, it is of utter importance to keep the spatial integrity of the features as much as possible. Cropping may introduce discontinuities along edges and misalignment between adjacent patches, and this is harmful for the dense volume segmentation task. Therefore, Z-Net chooses to only crop along the Z axis, instead of cropping along all three dimensions like Patch-64 and Patch-128 for traditional 3D DCNNs.

Z-Net also helps mitigating the class-imbalance problems, which is common in Patch-128 and Patch-64, especially for segmentation of small organs. Take aortic CT data as an example, in which only voxels around the center are labelled as the foreground. Patch-64 and Patch-128 cropped successively from the original volume will result in a large portion of sub-volumes cropped from the border regions having no foreground at all. If the patches are sampled selectively, the model might produce more false positives along the border regions. Z-Net ensures the presence of both foreground and background in all sub-volumes.

\subsection{Experimental Setup}
\label{sec:validation}

\subsubsection{Data collection}

20 aortic CT volumes from VISCERAL dataset \cite{jimenez2016cloud} are used in our experiment. All 20 volumes were randomly shuffled and divided into two groups for 2-fold cross validation. Each group contains 10 volumes for training, 2 volumes for evaluation and 8 volumes for testing. For the evaluation of the proposed Z-Net, Patch-512 of size $512 \times 512 \times 8$ was sampled, with a stride of 1 along the Z axis for training and a stride of 8 for the evaluation and testing. In order to compare with the baseline models, including 3D U-Net and V-Net, Patch-128 of size $128 \times 128 \times 64$ was generated, with strides of 128, 128, 8 along the X-, Y- and Z- axes for training and a stride of 128, 128, 64 for the evaluation and testing. In order to obtain enough training samples, 90\textdegree, 180\textdegree and 270\textdegree rotations about Z axis was applied for data augmentation. The maximum intensity value of the CT volume for each patient was used to normalize the CT volume intensity within $[0,1]$.

20 liver CT volumes from the SLiver07 \cite{heimann2009comparison} dataset were also used for the validation. All pre-processing procedures are the same as that for the VISCERAL dataset.

60 lung CT volumes from the Lung CT Segmentation Challenge 2017 \cite{yang2018autosegmentation} were used for the validation as well. We followed the instructions from the organizer and divided the 60 CT volumes into 36 and 24 volumes for the training and testing respectively. Hence 2-fold cross validation was not used for this dataset. A single 180\textdegree rotation was used for data augmentation. All other pre-processing procedures were the same as that for the VISCERAL dataset.

\subsubsection{Training configurations}

For all the training processes, Stochastic Gradient Descent (SGD) with a momentum of 0.9 was used as the default optimizer. The activation function for all 3D DCNNs was set as ReLU for consistency, even though originally Parametric ReLU was used in V-Net \cite{milletari2016v}. Weights were initialized with a truncated normal distribution, and biases were initialized as 0.1. Four initial learning rates (0.1, 0.05, 0.01, 0.005) were tested, and the one that achieved the highest segmentation accuracy was selected as the final result. The learning rate was dropped by half after the first epoch, and was further divided by 10 if the training process was longer than 4 epochs. 6 epochs were trained for the aortic data, while 4 epochs were trained for the liver and lung data. IoU, also known as the Jaccard Index, served as the metric for evaluating the performance of the segmentation:
\begin{equation}
    \text{IoU}=\frac{|Y \cap P|}{| Y \cup P|}
\end{equation}
where $Y$ is the ground truth and $P$ is the binarized prediction result. Foreground IoU was used to evaluate the segmentation accuracy. The prediction from the network was encoded in the one-hot fashion and cross-entropy loss was used, which can be calculated as:
\begin{equation}
    \xi(y,p)=-(y\log(p)+(1-y)\log(1-p))
\end{equation}
for binary-classification tasks, where $y$ is the ground truth value and $p$ is the prediction value for the segmentation given by the softmax function. All networks were trained using a CPU of Intel Xeon\textsuperscript{\textregistered} E5-1650 v4@3.60GHz$\times$12 and a GPU of Nvidia Titan Xp. The implementations of all the networks were based on TensorFlow.

\begin{table*}
\centering
\caption{The IoU results for 3D U-Net (Vanilla) and V-Net (Vanilla) on the training data generated by Patch-128 data generation, compared with the results for ZU-Net v1, ZU-Net v2, ZV-Net v1, ZV-Net v2 on the training data generated by Patch-512 data generation. The highest values are highlighted in bold. The accuracy improvement is calculated in (+ )}
\begin{tabular}{|l|l|l|l|l|l|l|l|} 
\hline
\multirow{2}{*}{DCNN} & \multirow{2}{*}{Patch} & Dataset & \multicolumn{2}{l|}{Aorta} & \multicolumn{2}{l|}{Liver} & Lung  \\ 
\cline{3-8}
& & Cross Validation & Fold 1 & Fold 2 & Fold 1 & Fold 2 & - \\ \hline
\multirow{3}{*}{U-Net} & Patch-128 & 3D U-Net (Vanilla) & 0.514 & 0.583 & 0.752 & 0.598 & 0.865 \\ 
\cline{2-8}
& \multirow{2}{*}{Patch-512} &ZU-Net v1 & 0.630(+0.116) & 0.652(+0.069) & \textbf{0.791}(+0.039) & 0.630(+0.032) & 0.868(+0.003) \\
\cline{3-8}
& & ZU-Net v2 & \textbf{0.640}(+0.126) & \textbf{0.655}(+0.072) & 0.783(+0.031) & \textbf{0.652}(+0.054) & \textbf{0.869}(+0.004) \\\hline
\cline{3-8}
\multirow{3}{*}{V-Net} & Patch-128 & V-Net (Vanilla) & 0.555 & 0.584 & 0.790 & 0.728 & 0.871 \\ 
\cline{2-8}
& \multirow{2}{*}{Patch-512} & ZV-Net v1 & \textbf{0.674}(+0.119) & 0.650(+0.066) & \textbf{0.852}(+0.062) & 0.756(+0.028) & \textbf{0.877}(+0.006) \\
\cline{3-8}
& & ZV-Net v2 & \textbf{0.674}(+0.119) & \textbf{0.660}(+0.076) & 0.851(+0.061) & \textbf{0.770}(+0.042) & 0.876(+0.005) \\ \hline
\end{tabular}
\label{tab:patch512}
\end{table*}

\begin{table*}
\centering
\caption{Comparison between the 3D U-Net (Vanilla), V-Net (three down-sampling layers), ZU-Net v1, ZV-Net v1, ZU-Net v2 and ZV-Net v2 on the aorta, liver and lung training data generated by Patch-512 method. The highest values are highlighted in bold.}
\begin{tabular}{|l|l|l|l|l|l|l|l|}
\hline
Dataset & \multicolumn{2}{l|}{Aorta} & \multicolumn{2}{l|}{Liver} & Lung & \multirow{2}{*}{Parameters} & \multirow{2}{*}{\begin{tabular}[c]{@{}l@{}}Training Time\\ (per 100 iterations)\end{tabular}} \\ \cline{1-6}
Cross Validation & Fold 1 & Fold 2 & Fold 1 & Fold 2 & - & &  \\ \hline
3D U-Net (Vanilla) & 0.616 & 0.614 & 0.764 & 0.613 & 0.866 & $1.40\times10^5$ & 70.36s \\ \hline
ZU-Net v1 & 0.630 & 0.652 & \textbf{0.791} & 0.630 & 0.868 &  $6.19\times10^4$ & 67.97s \\ \hline
ZU-Net v2 & \textbf{0.640} & \textbf{0.655} & 0.783 & \textbf{0.652} & \textbf{0.869} & \textbf{$5.73\times10^4$} & 67.02s \\ \hline
V-Net (three down-sampling layers) & 0.632 & 0.625 & 0.827 & 0.736 & 0.873 & $1.54\times10^7$ & 157.4s \\ \hline
ZV-Net v1 & \textbf{0.674} & 0.650 & \textbf{0.852} & 0.756 & \textbf{0.877} & $3.51\times10^6$ & 82.08s \\ \hline
ZV-Net v2 & \textbf{0.674} & \textbf{0.660} & 0.851 & \textbf{0.770} & 0.876 & \textbf{$3.33\times10^6$} & 80.11s \\ \hline
\end{tabular}
\label{tab:result-znet}
\end{table*}

\begin{figure*}
    \centering
    \framebox{\includegraphics[width=0.7\textwidth]{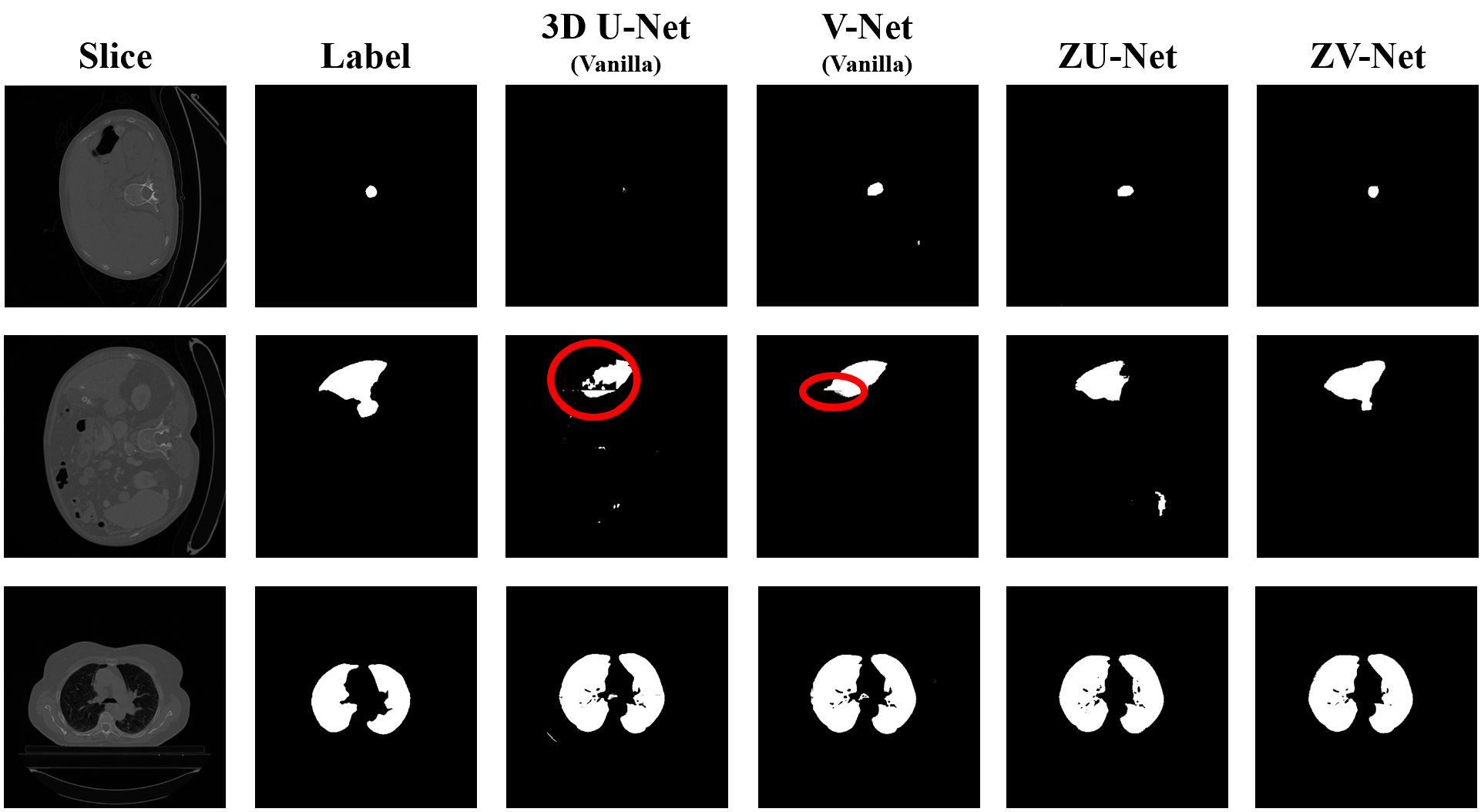}}
    \caption{One 2D slice from the CT volume of a randomly selected patient is shown with the ground truth, the segmentation result of training 3D U-Net (Vanilla) and V-Net (Vanilla) on patch-128 data and of training ZU-Net v2 and ZV-Net v2 on Patch-512 data. The regions in the red circles show the misalignment in prediction results in traditional 3D DCNNs.}
    \label{fig:result-slices}
\end{figure*}

\section{Result}
\label{sec:result}

In order to compare the proposed Z-Net framework with traditional 3D DCNNs, the vanilla version of 3D U-Net and V-Net were trained using the training data generated with Patch-128 data generation, which served as the baseline. Patch-64 was not compared in our experiment due to the convergence difficulty with the aortic dataset, which was as expected. Even multiple pre-processing methods have been used to optimize the data, a heavy class-imbalance was still presented using Patch-64 data generation. The failure of training 3D DCNNs on Patch-64 data also supports our statement of the three deficiencies in Sec. \ref{sec: Intro}, which is also the motivation for proposing Z-Net. ZU-Net v1, ZV-Net v1, ZU-Net v2 and ZV-Net v2 were trained on the training data generated with Patch-512 data generation. Detailed results are stated in Sec. \ref{sec:result-patch512}.

In order to do ablation analysis and figure out where the accuracy improvement of the proposed Z-Net comes from, we further test 3D U-Net and V-Net on the same Patch-512 data used in Z-Net. As the training patches generated from Patch-512 are only with a size of 8 in the Z axis, hence 3D U-Net and V-Net with three down-sampling layers are trained. Detailed results are illustrated in Sec. \ref{sec:result-znet}. CT volumes and 2D slices of patients are randomly selected to show the detailed segmentation difference between different methods in Sec. \ref{sec:result-exmaple} while the training loss curves for different methods are shown in Sec. \ref{sec:result-convergence}.  

\subsection{Z-Net}
\label{sec:result-patch512}

The mean segmentation IoUs of training vanilla 3D U-Net and V-Net on the aorta, liver and lung training data generated from Patch-128 data generation, and the mean segmentation IoUs of training ZU-Net v1, ZU-Net v2, ZV-Net v1 and ZV-Net v2 on the aorta, liver and lung training data generated by Patch-512 data generation are shown in Tab. \ref{tab:patch512}. It can be observed that the proposed Z-Net achieves $6.6\%-12.6\%$, $2.8\%-6.2\%$, and $0.3\%-0.6\%$ mean IoU improvements on the aorta, liver and lung data respectively, over the 3D U-Net and V-Net.

One observation is that the order of IoU improvement is aorta $>$ liver $>$ lung, which is in the reverse order of their physical size. This result indicates that large organs such as lung and liver appear to be less affected by the shortages of existing 3D DCNNs, whereas small organs such as aorta gain more improvement from the proposed Z-Net. An up to $12.6\%$ IoU improvement proves the severe issue of class-imbalance presented by traditional 3D DCNNs.

\subsection{Ablation analysis}
\label{sec:result-znet}

The mean segmentation IoUs of 3D U-Net and V-Net with three down-sampling layers, as well as ZU-Net v1, ZU-Net v2, ZV-Net v1 and ZV-Net v2 on the aorta, liver and lung  Patch-512 training data are shown in Tab. \ref{tab:result-znet}. It can be observed that for the aorta and liver segmentation, the ZU-Net and ZV-Net in both modes outperform the 3D U-Net and V-Net by around $3\%$. On the other hand, the segmentation IoU improvement for the lung segmentation is smaller, being $0.1\%$ and $0.3\%$. It can be concluded that the proposed Z-Net out-performs all baselines in all validations, even training all baselines on the same Patch-512 data. It can also be concluded that mode 2 outperforms mode 1 in most tests, except cross validation 1 for the liver segmentation. These conclusions prove the motivation of Z-Net stated in Sec. \ref{sec: Intro}, where traditional 3D DCNNs crop the original and large CT volume into small and isotropic patches is less optimal for medical volume segmentation.

The number of trainable parameters and the training time for 3D U-Net (Vanilla), V-Net with three down-sampling layers, ZU-Net v1, ZU-Net v2, ZV-Net v1 and ZV-Net v2 are also shown in Tab. \ref{tab:result-znet}. It can be seen that the proposed Z-Net variants possess significantly less trainable parameters. The modified versions of ZU-Net contain less than a half number of trainable parameters than the 3D U-Net (Vanilla), and the number of parameters for ZV-Net is only around $\frac{1}{5}$ of the V-Net with three down-sampling layers. The table also shows faster training speed after the modifications of the networks resulting from fewer trainable parameters, especially for V-Net. The training speed is measured as the average amount of time in seconds for the networks to train 100 iterations.

\subsection{Segmentation examples}
\label{sec:result-exmaple}

\begin{figure}
    \centering
    \framebox{\includegraphics[width=0.47\textwidth]{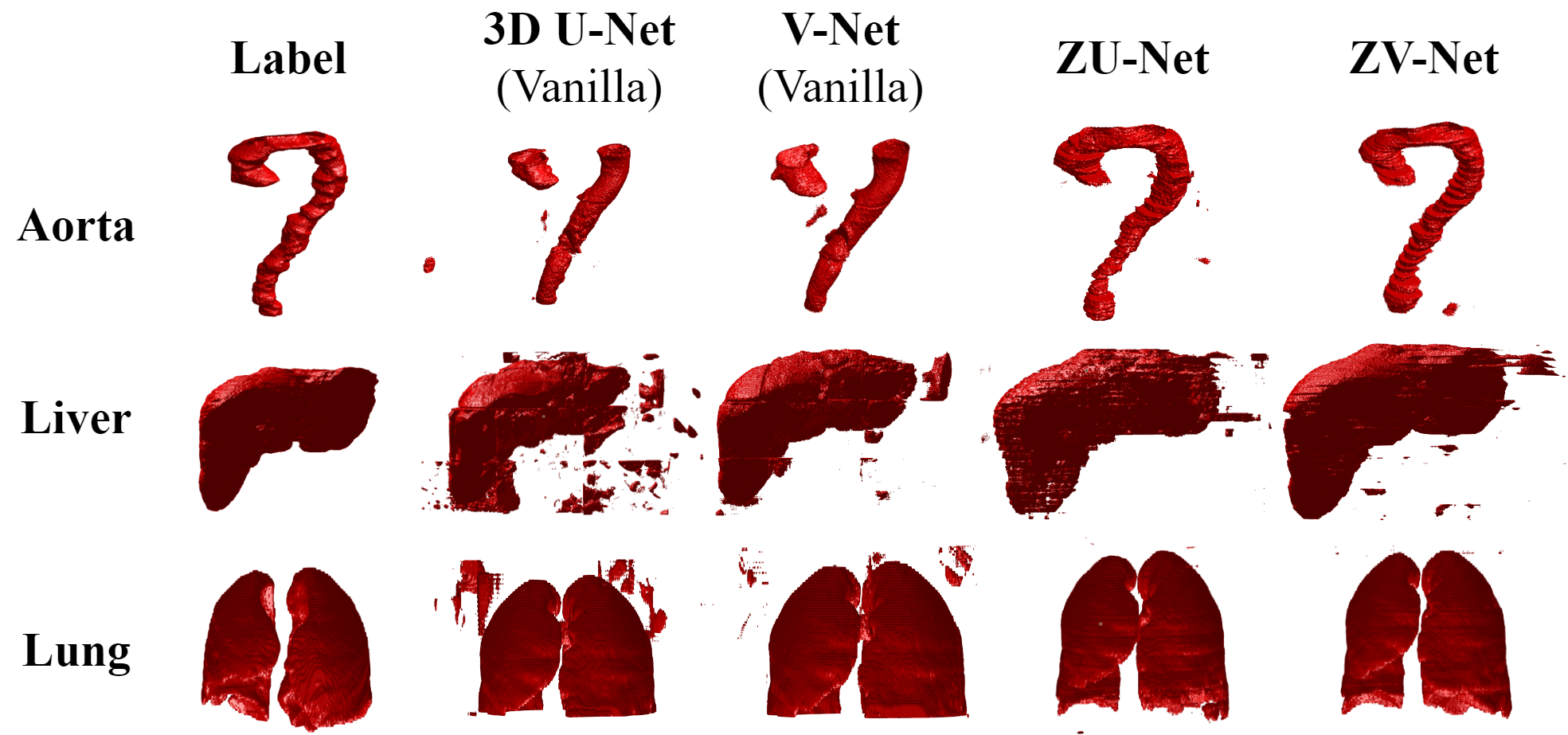}}
    \caption{One patient is randomly selected for visualizing the detailed volume segmentation result of vanilla 3D U-Net and vanilla V-Net on the aorta, liver and lung Patch-128 training data and with training ZU-Net v2 and ZV-Net v2 on the aorta, liver and lung Patch-512 training data.}
    \label{fig:result-volume}
\end{figure}

One patient was selected to show the detailed 3D segmentation result of the aorta, liver and lung, with the 3D U-Net (Vanilla) and V-Net (Vanilla) training on Patch-128 data, and with the proposed ZU-Net and ZV-Net under mode 2 training on Patch-512 data, as shown in Fig. \ref{fig:result-volume}. We can see that the proposed method in this paper achieved noticeable better segmentation result with much less false positives and noises.

For better visualization, we also show some detailed 2D slice segmentation results of the aorta, liver and lung with different methods in Fig. \ref{fig:result-slices}. It is also obvious that the proposed method in this paper achieved visually better segmentation results without misalignment between patches along the X and Y axes.

\subsection{Convergence}
\label{sec:result-convergence}

\begin{figure}
    \centering
    \framebox{\includegraphics[width=0.4\textwidth]{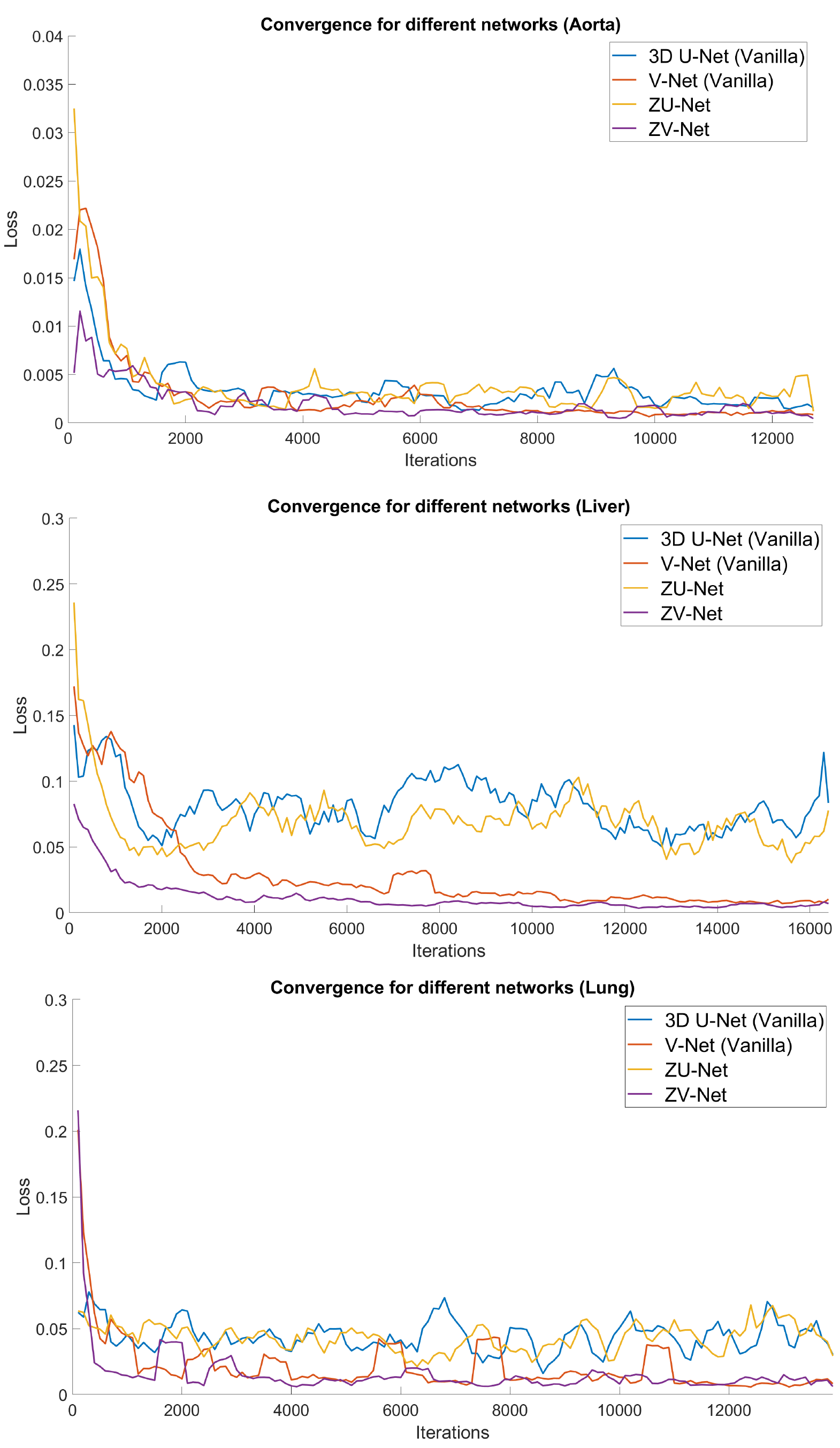}}
    \caption{The loss plots for training vanilla U-Net, vanilla V-Net and ZU-Net v2, ZV-Net v2 on the aorta, liver and lung dataset.}
    \label{fig:result-loss}
\end{figure}

The loss during the training of vanilla 3D U-Net, vanilla V-Net, ZU-Net v2 and ZV-Net v2 for the aorta, liver and lung segmentation are shown in Fig. \ref{fig:result-loss}, which illustrates that the proposed Z-Net achieved lower losses for the three datasets than the vanilla 3D U-Net and vanilla V-Net, but the convergence speed is similar.

\section{Discussion}
\label{sec:discussion}

The proposed Z-Net maintains a full field-of-view in the XY slices while maintains as many slices as possible in the Z axis according to the GPU memory. The effect of cropping along all X, Y and Z axis and then assembling the prediction result back can be seen clearly in Fig. \ref{fig:result-slices}, where discontinuities between adjacent predictions result in the large gaps and holes. Z-Net retains the spatial integrity of features along the XY plane, giving a better result observed from 2D slices. Furthermore, the Z-Net not only compensates the class-imbalance issue caused by Patch-128 and Patch-64 in traditional 3D DCNNs, but also augments the number of training data and keeps the GPU memory under an affordable value. Promising improvements on the segmentation accuracy, especially for small targets like the aorta, can be observed from the results, demonstrating the effectiveness of the proposed Z-Net method. 

Two modes of Z-Net were also explored, either with or without the intermediate 1D convolutions between 2D convolutions. According to the validation results, mode 2 slightly out-performed mode 1, and the number of parameters and training time are also reduced. This indicates that a single 1D convolution before each down-sampling or up-sampling layer is sufficient for inter-slice context extraction.

Overall, the total improvements in IoU of up to  $12.6\%$, $6.2\%$ and $0.6\%$ were achieved for the aortic, liver and lung CT volume segmentation respectively with the proposed Z-Net framework, compared to the original 3D U-Net and V-Net. The segmented 3D shape in this paper is very useful for many advanced medical tasks, i.e. 3D shape instantiation and registration for 3D navigation in robot-assisted MIS.

\section{Conclusion}
\label{sec:conclusion}

To summarize, Z-Net is proposed to extract the information in the XY-planes and Z-axis separately for training with $512\times512\times8$ patches. It can alleviate the issues for traditional 3D DCNNs, including class-imbalance, discontinued segmentation results and limited field-of-view for each individual volume, hence the current volume segmentation accuracy and speed can be increased noticeably. The medical CT volume segmentation in this paper both automates and supplies an essential pre-operative knowledge for achieving intra-operative 3D navigation for robot-assisted MISs. In the future, we will work on deeper network designs and better kernel decomposition methods for medical volume segmentation to further improve the segmentation performance.

\bibliographystyle{IEEEtran}
\bibliography{ICRA2020.bib}

\end{document}